\documentclass[10pt,a4paper,twocolumn, superscriptaddress]{revtex4-2}
\usepackage[latin1]{inputenc}
\usepackage{amsmath}
\usepackage{amsfonts}
\usepackage{amssymb}
\usepackage{graphicx}
\graphicspath{{../Figures/}}
\usepackage[separate-uncertainty=true]{siunitx}
\usepackage{xcolor}
\usepackage{float}
\usepackage{mathtools}
\usepackage{hyperref}

\DeclarePairedDelimiterX\braket[2]{\langle}{\rangle}{#1 \delimsize\vert #2}

\begin{document}
\author{R. St\"uhler}
\email{raul.stuehler@physik.uni-wuerzburg.de}
\author{A. Kowalewski}
\author{F. Reis}
\affiliation{Physikalisches Institut and W\"urzburg-Dresden Cluster of Excellence ct.qmat, Universit\"at W\"urzburg, D-97074 W\"urzburg, Germany}
\author{D. Jungblut}
\affiliation{Institut f\"ur Theoretische Physik und Astrophysik and W\"urzburg-Dresden Cluster of Excellence ct.qmat, Universit\"at W\"urzburg, D-97074 W\"urzburg, Germany}
\author{F. Dominguez}
\affiliation{Institut f\"ur Theoretische Physik und Astrophysik and W\"urzburg-Dresden Cluster of Excellence ct.qmat, Universit\"at W\"urzburg, D-97074 W\"urzburg, Germany}
\affiliation{Institute for Mathematical Physics, TU Braunschweig, 38106 Braunschweig, Germany}
\author{B. Scharf}
\affiliation{Institut f\"ur Theoretische Physik und Astrophysik and W\"urzburg-Dresden Cluster of Excellence ct.qmat, Universit\"at W\"urzburg, D-97074 W\"urzburg, Germany}
\author{G. Li}
\affiliation{Institut f\"ur Theoretische Physik und Astrophysik and W\"urzburg-Dresden Cluster of Excellence ct.qmat, Universit\"at W\"urzburg, D-97074 W\"urzburg, Germany}
\affiliation{School of Physical Science and Technology, ShanghaiTech University, Shanghai 201210, China}
\affiliation{ShanghaiTech Laboratory for Topological Physics, Shanghai 200031, China}
\author{J. Sch\"afer}
\affiliation{Physikalisches Institut and W\"urzburg-Dresden Cluster of Excellence ct.qmat, Universit\"at W\"urzburg, D-97074 W\"urzburg, Germany}
\author{E. M. Hankiewicz}
\affiliation{Institut f\"ur Theoretische Physik und Astrophysik and W\"urzburg-Dresden Cluster of Excellence ct.qmat, Universit\"at W\"urzburg, D-97074 W\"urzburg, Germany}
\author{R. Claessen}
\affiliation{Physikalisches Institut and W\"urzburg-Dresden Cluster of Excellence ct.qmat, Universit\"at W\"urzburg, D-97074 W\"urzburg, Germany}

\date{\today}
\title{Lifting topological protection in a quantum spin Hall insulator by edge coupling}

\maketitle
\textbf{
The scientific interest in two-dimensional topological insulators (2D TIs) is
currently shifting from a more fundamental perspective to the exploration and
design of novel functionalities. Key concepts for the use of 2D TIs in
spintronics are based on the topological protection and spin-momentum locking of
their helical edge states. In this study we present experimental evidence that
topological protection can be (partially) lifted by pairwise coupling of 2D TI
edges in close proximity. Using direct wave function mapping via scanning
tunneling microscopy/spectroscopy (STM/STS) we compare isolated and coupled
topological edges in the 2D TI bismuthene. The latter situation is realized by
natural lattice line defects and reveals distinct quasi-particle interference
(QPI) patterns, identified as electronic Fabry-P\'{e}rot resonator modes. In
contrast, free edges show no sign of any single-particle backscattering. These
results pave the way for novel device concepts based on active control of
topological protection through inter-edge hybridization for, e.g., electronic
Fabry-P\'{e}rot interferometry.
}\par


\begin{figure*}
	\centering
	\includegraphics[width = 0.95\textwidth ]{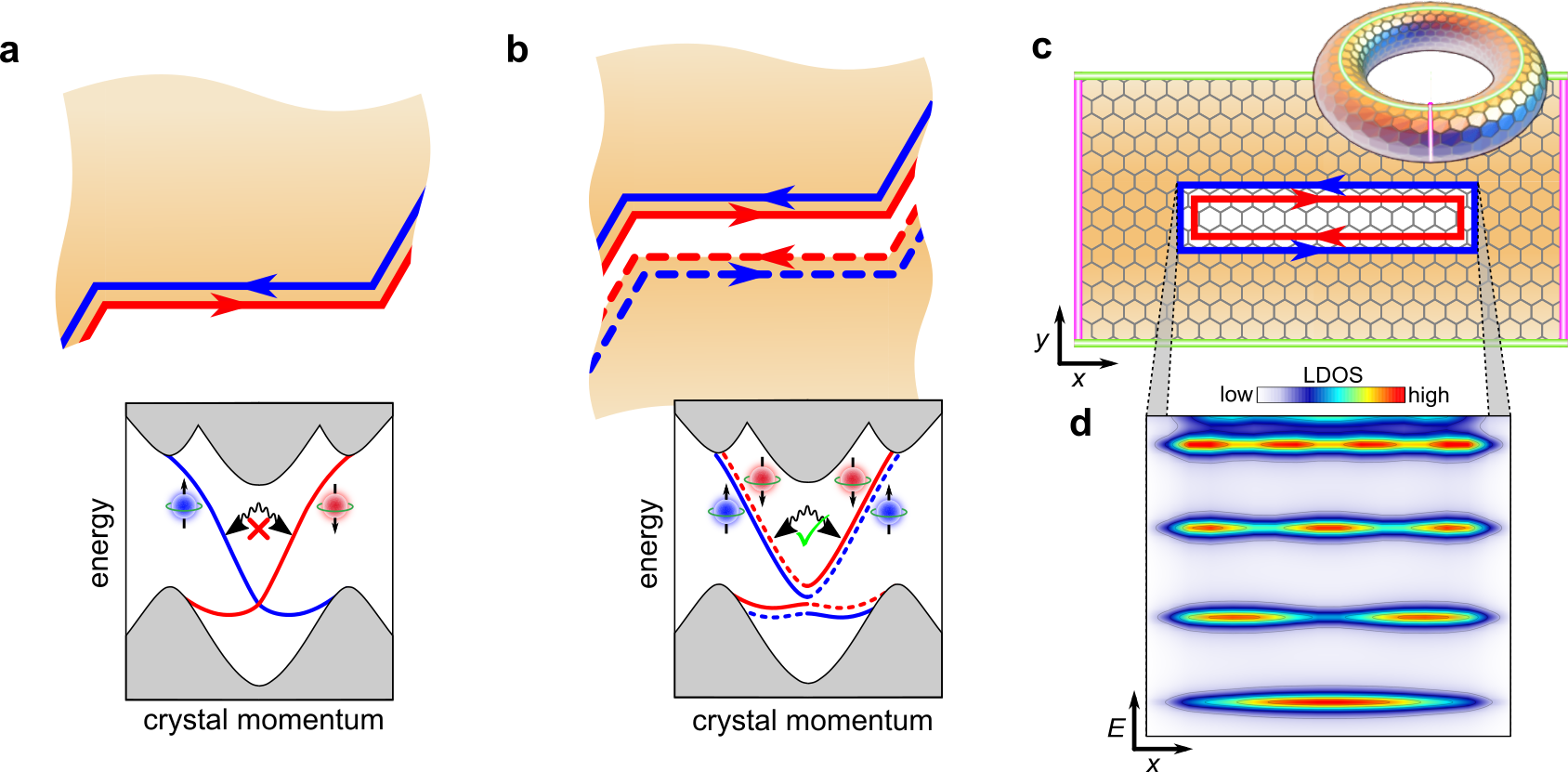}
	\caption{
\textbf{Free vs. coupled helical edge states.}
\textbf{a}, 
Edge segment of a generic 2D TI material. A pair of helical edge states
is bound to the edge. Back-scattering from defects such as the edge kinks is impeded by
topologically protected spin-momentum locking, i.e., left-moving spin-up states
(blue) cannot scatter with right-moving spin-down states (red). The
corresponding one-dimensional band structure is schematically depicted below.
\textbf{b}, 
Two opposing edge segments in close proximity. Spatial overlap of both pairs of
helical edge states induces hybridization (or tunneling) between both edges,
thereby allowing inter-edge scattering from one segment to the other across the
boundary. Strikingly, this opens a channel for back scattering from the edge
kinks, i.e., the topological protection becomes partially lifted due to
inter-edge hybridization. The corresponding band structure is depicted below.
\textbf{c},
Tight-binding model of coupled edge states in the 2D TI bismuthene. The white
area indicates a topologically trivial line defect along the $x$-direction
embedded in the topological bulk (light brown). The green and pink boundaries of
the finite-size bulk are connected via periodic boundary conditions, resulting
in the depicted torus. The trivial line defect connects two opposite edges; the
resulting wave function overlap between both sides induces hybridization of
like-spin edge states across the line defect ($y$-direction) as in \textbf{b}.
\textbf{d}, 
Energy-dependent LDOS of the tight-binding model along the line defect in
\textbf{c} (integrated across its $y$-width). The line defect induces standing
wave excitations bound to its longitudinal extent (along $x$) and, as the
underlying helical states of bismuthene exhibit an approximate linear
dispersion, are linearly quantized in energy $E$. For details of the calculation
see the Supplementary Information.
	}
	\label{fig:1}
\end{figure*}

In 2D TIs the bulk-boundary correspondence enforces the existence of metallic
gap states confined to the one-dimensional (1D) edges of the material. These
edge states are spin-polarized and have their spin rigidly locked to the
electron's momentum~\cite{Hasan2010}. As a consequence electrons moving along
the edge cannot be backscattered by any non-magnetic defects, as schematically
depicted in Fig.~\ref{fig:1}a. This topological edge state protection lies at
the heart of the celebrated quantum spin Hall (QSH) effect~\cite{Koenig2007}. It
is obvious that the unique properties of the topological edge states lend
themselves to a plethora of novel functionalities and application ideas, ranging
from low-power consumption electronics to innovative spintronics devices to
possible solid-state realizations of qubits ~\cite{Qi2011}.

Most theoretical device concepts are based on direct control of the topological
edge states by external stimuli, such as electric or magnetic fields, or by
bringing them into spatial proximity to ferromagnetic or superconducting
materials. Ideas include in particular various types of field-effect transistors
(FET). For example, the symmetry-breaking effect of an electric gate field can
be used to trigger a phase transition of the 2D TI to a trivial insulator, as
recently demonstrated for Na$_3$Bi~\cite{Collins2018}. The on/off
characteristics of such a FET is determined by the complete quenching of the
current-carrying topological edge states~\cite{Liu2013,Michetti2013,Qian2014}.
In an alternative FET concept relying on much smaller gate fields the Fermi
level is toggled between an in-gap position and the bulk band edges. In the
former situation the edge states carry a dissipationless current, while in the
latter they will couple to the dissipative bulk states, thereby effectively
losing their topological protection. The resulting promotion of backscattering
from impurities and phonons is estimated to allow on/off ratios of more than two
orders of magnitude~\cite{Vandenberghe2017}.

%

A more direct way of controlling and eventually lifting topological protection
is achieved by tunneling between opposite edges of a 2D TI
~\cite{Zhou2008,Krueckl2011,Sternativo2014, Romeo2012, Maciel2021, Ishida2020,
Takagaki2012, Takagaki2015}. The resulting hybridization between
right(left)-moving electrons on one edge with their left(right)-moving partners
of like spin on the opposite edge will open a small gap at the Dirac point
and--even more importantly--create a channel for electron backscattering without the need to break time-reversal symmetry, as
depicted in Fig.~\ref{fig:1}b. A possible realization of such a situation is a 
narrow constriction in a 2D TI as, e.g., engineered by nanopatterning or defined by 
suitable line defects in the atomic lattice~\cite{Lima2016}.
Interedge tunneling will then lead to the formation of Fabry-P\'{e}rot-type
resonances along the constriction and hence a modulated transmission through the
device, controlled by the position of the Fermi level~\cite{Sternativo2014}.
While numerous theoretical proposals along this line have been put forward,
there exist surprisingly few experimental studies on edge coupling in a QSH
insulator. Strunz \textit{et al.}~\cite{Strunz2020} have recently studied the
effect of Coulomb interaction between 2D TI edges in spatial proximity, whereas
Jung \textit{et al.}~\cite{Jung2021} focus on the tunneling-induced gap opening
in the 1D edge states of a topological crystalline insulator.


In this paper we examine the effect of topological edge coupling by spatial
mapping of the resulting wave function, thereby revealing direct experimental
evidence for the loss of topological protection without breaking time-reversal symmetry. For this purpose we use
bismuthene as prototypical 2D TI~\cite{Reis2017}, in which narrow constrictions
appear naturally in the form of line defects. Bismuthene consists of a honeycomb
monolayer of Bi atoms covalently bonded to a SiC(0001) substrate. Its band
structure features a large topological band gap of $\sim \SI{0.8}{\electronvolt}$~\cite{Maklar2021}, inducing helical edge states which due to their narrow spatial confinement of only 
a few Angstroms are ideally suited for wave function mapping by
STM/STS~\cite{Reis2017,Stuehler2020}. 

First insight into the physics of edge coupling in bismuthene can be obtained
from a tight-binding (TB) approach based on first-principles
calculations~\cite{Reis2017,Li2018} (see Supplementary Information for details). In
our model we place a line defect into the bismuthene bulk, consisting of a
stripe-like region of topologically trivial material, as depicted in
Fig.~\ref{fig:1}c. Effectively this can be thought of as two finite-length edges
of the topological bulk coupled across a narrow trivial gap. Indeed, we observe the 
formation of helical edge states along both edges of the line defect which 
counterpropagate with opposite helicity. Importantly, the line defect is narrow 
enough to generate a spatial overlap of the edge state wave functions across 
the defect (i.e., in $y$-direction), resulting in the expected hybridization of 
like spin states. This in turn allows backscattering at the two terminations of the line defect
and subsequently induces the formation of standing waves, similarly to the behavior of light
waves in an optical resonator~\cite{Liang2001, Seo2010}. Figure~\ref{fig:1}d
displays the corresponding Fabry-P\'{e}rot oscillations in the local density of
states (LDOS). The quantized resonance levels are linearly spaced in energy as a
direct consequence of the linear dispersion relation of the underlying bismuthene
edge states~\cite{Reis2017}.


\begin{figure*}
	\centering
	\includegraphics[width = 0.95\textwidth ]{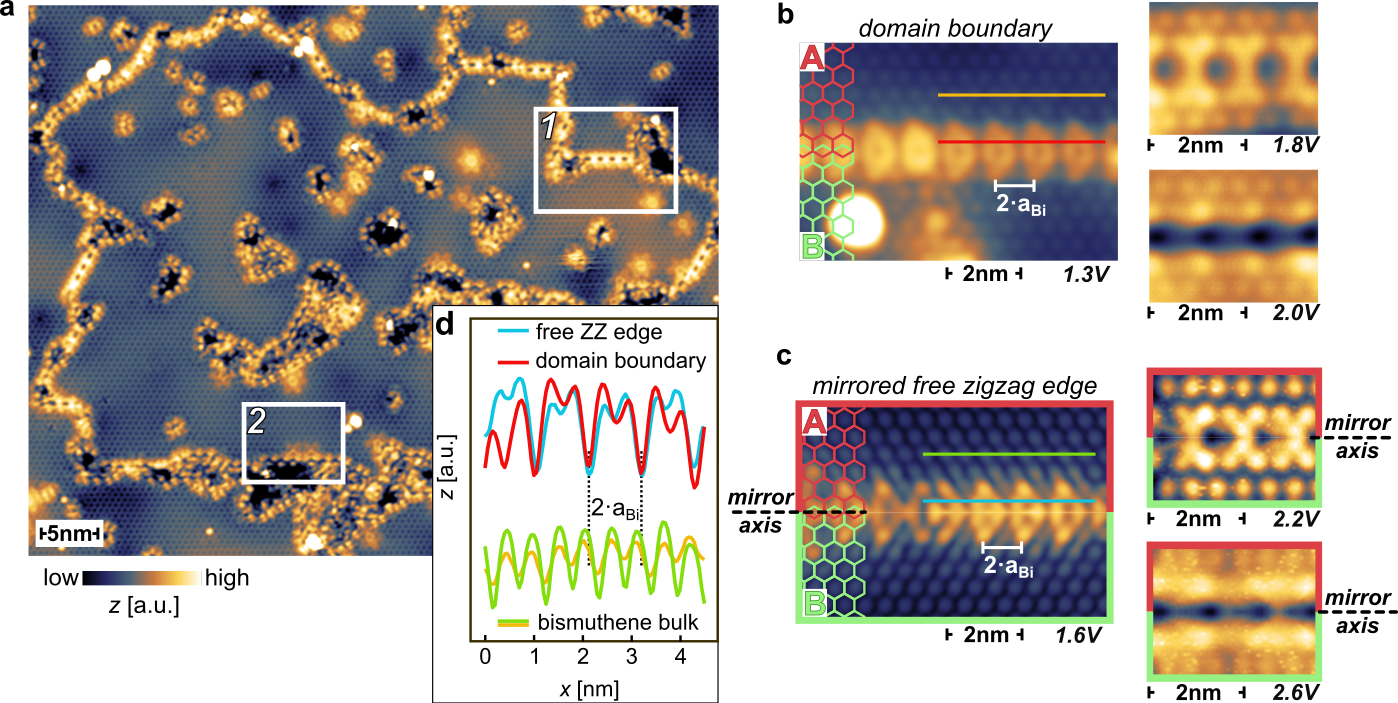}
	\caption{
\textbf{Domain boundary and zigzag edge topography.}
\textbf{a}, 
STM constant-current map showing a topographic overview of
bismuthene. Its $\sqrt{3}\times \sqrt{3}R30^\circ$ superstructure
with respect to the SiC(0001) substrate surface results in the 
formation of three distinct phase domains. The meandering network 
of line defects are DBs separating the domains from each other. 
Window 1  marks a particularly well-ordered straight DB section. 
The map also shows free zigzag edges encircling uncovered 
substrate surface due to incomplete monolayer growth, see window 2.
$V_{\text{set}} = \SI{-1.0}{\volt}$, $I_{\text{set}} = \SI{10}{\pico\ampere}$.
\textbf{b}, 
STM topography of a DB recorded at different set-points ($V_{\text{set}}= \left[
\SI{1.3}{\volt}, \SI{1.8}{\volt}, \SI{2.0}{\volt} \right]$), revealing
bias-dependent structural features. $ I_{\text{set}}= \left[\SI{50}{\pico
\ampere}, \SI{50}{\pico \ampere}, \SI{50}{\pico \ampere} \right]$. 
\textbf{c},
STM topography of a free zigzag edge recorded at $V_{\text{set}}=
\left[ \SI{1.6}{\volt}, \SI{2.2}{\volt}, \SI{2.6}{\volt} \right] $. The images
show the free zigzag edge mirrored at the marked mirror axis such that domain
'A' is mirrored into domain 'B' to properly generate an adsorption site shift to
mimic the situation at a DB. The registry mismatch between both domains can 
be seen in the respective hexagonal lattices. 
$I_{\text{set}}= \left[\SI{70}{\pico \ampere}, \SI{100}{\pico \ampere} ,
\SI{100}{\pico \ampere} \right]$. The procedure of mirroring is exemplified for
one zigzag edge in Fig.~S5 in the Supplementary Information. 
\textbf{d}, 
STM constant-current line profiles from the STM topographies 
in \textbf{b} and \textbf{c}, highlighting the similarly doubled periodicities 
of free zigzag edge and DB with respect to the bismuthene bulk.
	}
	\label{fig:2}
\end{figure*}


For the experiments, the bismuthene monolayers were prepared by molecular beam
epitaxy using a Knudsen cell as Bi source and SiC(0001) as supporting substrate
(for details see Ref.~\cite{Reis2017} and the Method section). Bismuthene forms
a $\sqrt{3}\times \sqrt{3}R30^\circ$ superstructure of Bi atoms in planar
honeycomb geometry on the SiC(0001) substrate, clearly resolved in the
topographic STM overview of Fig.~\ref{fig:2}a. For such a superlattice three
distinct yet equivalent phase registries exist with respect to the substrate
lattice, resulting in the formation of phase domains separated by
phase-slip domain boundaries (DBs)~\cite{Lahiri2010}. In
Fig.~\ref{fig:2}a the DBs appear as meandering line defects and, importantly,
display a distinct periodic structure along their piecewise straight sections,
see window 1. The DBs are observed only between zigzag
terminations of the connected bismuthene domains, indicating that DBs formed by
armchair terminations are energetically unfavorable. While the individual DB
straight sections in Fig.~\ref{fig:2}a are relatively short we generally observe
a wide distribution of lenghts up to $\sim \SI{25}{\nano\metre}$ (see
Fig.~S3 in the Supplementary Information). In addition to the DBs we also
observe free edges with zigzag termination at regions where the film growth is
locally incomplete, see window 2 in Fig.~\ref{fig:2}a for an example. In this
case the terminated domain is not directly connected to another domain.


Further insight into the local DB structure can be obtained by direct comparison
to the free zigzag edges, using set-point dependent STM constant current
measurements. The DBs reveal characteristic topographic features at certain bias
voltages as shown in Fig.~\ref{fig:2}b: arrow-shaped segments at
\SI{1.3}{\volt}, bone-shaped segments at \SI{1.8}{\volt}, and a groove at
\SI{2.0}{\volt}. Fig.~\ref{fig:2}c demonstrates that the DB structure can indeed
be understood as the combination of two free zigzag edges: it is composed of
bias-dependent STM images of a free zigzag edge, additionally mirrored and
stitched together at the appropriate mirror axis. One can clearly identify the
same structural features as for the DB in Fig.~\ref{fig:2}b, though slightly
shifted in bias. This remarkable agreement allows a precise identification of
the lattice phase slip between the domains 'A' and 'B' joined by the DB,
visualized by the red and green honeycomb lattices in Figs.~\ref{fig:2}b and c.
A second result of this comparison is the determination of the separation
between the two zigzag edges forming the DB, which amounts to \SI{15}{\angstrom}
(see Fig.~S1 in the Supplementary Information). Finally, excellent correspondence is also observed for
the line profiles of both DB and free zigzag edge (Fig.~\ref{fig:2}d). In
particular, the profiles reveal a period doubling of DB and free
edge ($2a_{\text{Bi}} = \SI{10.7}{\angstrom}$) with respect to the lattice
constant $a_{\text{Bi}}$ of bulk bismuthene. This behavior is strongly
reminiscent of the so-called (57) reconstruction reported for graphene zigzag
edges~\cite{Girit2009, Koskinen2008,Koskinen2009}. Furthermore, for 
graphene/Ni(111) (558)-type DBs have been observed which form from  
two merging (57)-reconstructed zigzag edges~\cite{Lahiri2010}, in close 
correspondence to our present findings.

\begin{figure*}
	\centering
	\includegraphics[width = \textwidth ]{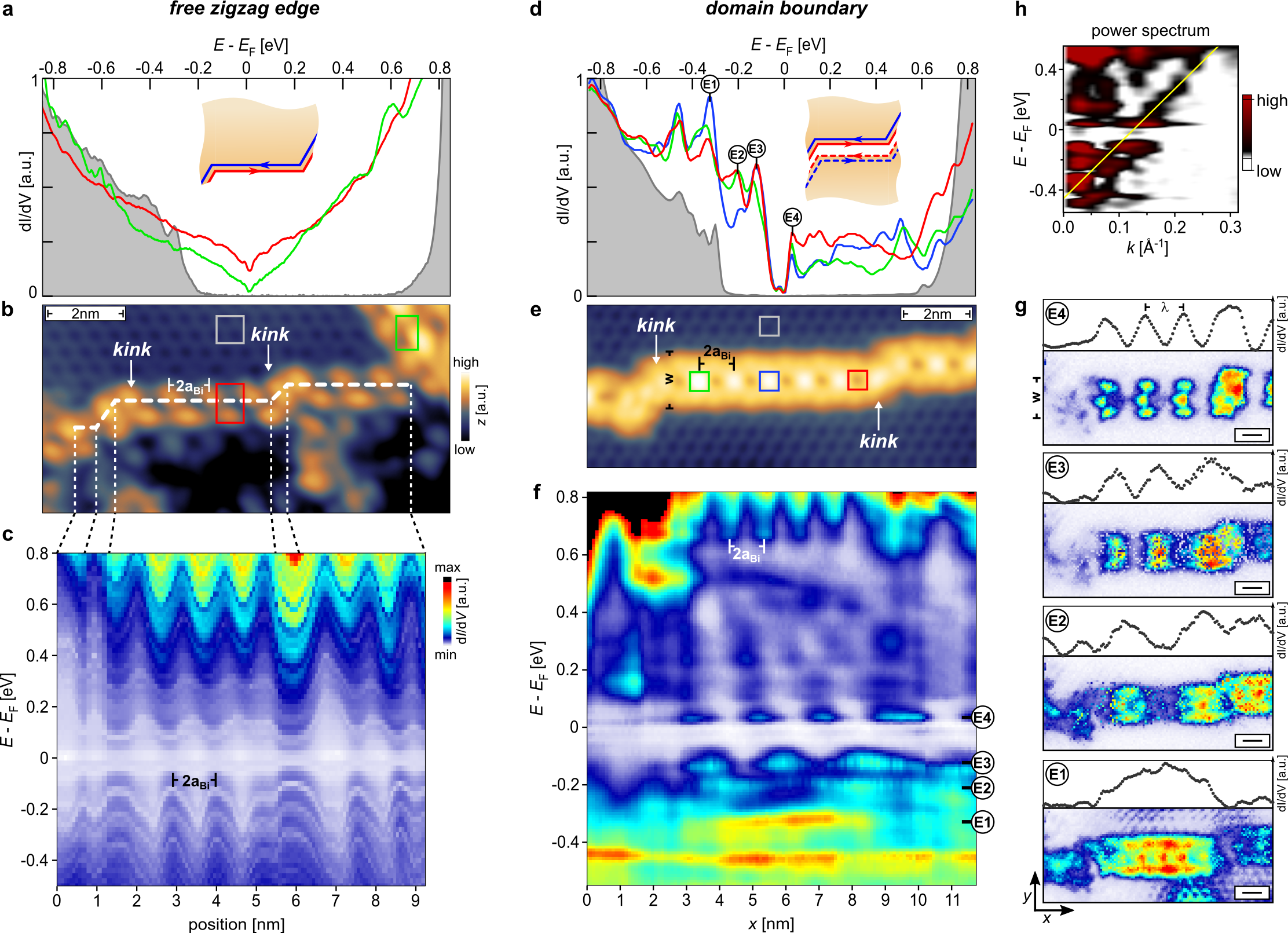}
	\caption{
\textbf{Elecronic properties of free zigzag edge states vs. domain boundary states.} 	
\textbf{a}, 
	$dI/dV$ tunneling spectra on and off a free bismuthene zigzag edge, measured by
	STS at the positions marked in \textbf{b}. While the bulk spectrum (gray) shows
	a clear insulating band gap, the $dI/dV$ spectra taken on the edge (red, green)
	verify the existence of a metallic edge channel smoothly filling the gap.
\textbf{b}, 
	Constant-current STM topography of a free zigzag edge ($V_{\text{set}}= \SI{-1.2}{\volt}$,
	$I_{\text{set}}= \SI{20}{\pico \ampere}$). The edge is divided into
	straight edge segments of variable length separated by kinks, as indicated by
	the white-dashed line. The colored squares mark the positions, where the spectra
	in \textbf{a} are taken.
\textbf{c}, 
	Energy-dependent $dI/dV$ signal versus position along the white-dashed path marked in \textbf{b}. The periodic
	modulation corresponds to the structural edge periodicity $2a_{\text{Bi}}$.
\textbf{d}, 
	$dI/dV$ spectra on and off a bismuthene DB, taken at the positions
	marked in \textbf{e}. The bulk spectrum (gray) agrees with the one in
	\textbf{a}. The spectra taken on the DB (green, blue, red) (nearly) fill the
	band gap and are characterized by the presence of intense peaks at
	discrete energies, in remarkable contrast to the zigzag edge spectra in
	\textbf{a}.
\textbf{e}, 
	STM topography of a DB ($V_{\text{set}}= \SI{-0.85}{\volt}$, $I_{\text{set}}=
	\SI{80}{\pico \ampere}$). Similar to the free zigzag edge, the DB consists of
	piecewise straight sections connected by kinks. The colored squares mark the
	positions, where the spectra in \textbf{d} are taken.
\textbf{f}, 
	Energy-dependent $dI/dV$ signal versus position along the DB in \textbf{b},
	integrated over the DB width $w$. Energy-dependent standing wave excitations are
	observed at the marked energies $E_1$ to $E_4$, additional to the structural $2a_{\text{Bi}}$ modulation.
\textbf{g}, 
	Two-dimensional $dI/dV$ maps of the DB in \textbf{e}, taken at the 
	discrete energies $E_1$ to $E_4$ defined in \textbf{f}. The
	corresponding $dI/dV$ profile along the DB (integrated over its width $w$) is shown above
	each map. The modulation wavelength $\lambda$ can be extracted from the $dI/dV$ profiles. The scale bar
	is \SI{1}{nm}.
\textbf{h}, 
	Power spectrum ($|\text{FFT}\left[dI/dV(E;x)\right](k)|^2$) of the $dI/dV$ map in
	\textbf{f}. The yellow line is a guide to the eye with a slope of $\SI{3.6}{\electronvolt\angstrom}$.
	}
	\label{fig:3}
\end{figure*}


Having thus established that the bismuthene DBs can structurally be viewed as coupled
zigzag edges we now turn to their electronic properties, again in direct
comparison to their free counterparts. Because the helical edge states in bismuthene
are confined to within only a few atomic distances from the boundary, their spatial
charge distribution is ideally addressed by STM/STS. Specifically, the local
differential conductivity ($dI/dV$) recorded by this technique is a direct
measure of the LDOS. In a perfect lattice the LDOS is spatially modulated only by the
lattice periodicity. However, local imperfections, such as the kinks in our DBs,
can give rise to additional modulation of the $dI/dV$ signal through
QPI, i.e., the interference between incoming and
elastically scattered electron wave. The modulation wavelength is given by
$\lambda(E)=\pi/k(E)$, where $E$ and $k$ are energy and momentum of the
electron, respectively. QPI is thus a powerful method to probe the presence vs.
absence of single particle backscattering, and hence of topological protection.


Experimentally, the existence of metallic edge states in bismuthene has so far
only been established for armchair edges induced at substrate terrace
steps~\cite{Reis2017, Stuehler2020}. It remains to be demonstrated that
topological edge states also exist at free zigzag edges. Figure~\ref{fig:3}b 
depicts such an edge, terminating the bismuthene domain in the
upper half of the image. The edge is not a continuously straight line, but
consists of several straight segments of variable length, interconnected by kinks
where the edge changes direction by multiples of 60\textdegree, thereby 
maintaining the zigzag nature of the edge. 

We first measure the $dI/dV$ signal far from the edge in the bismuthene bulk at
a position marked by the gray square in Fig.~\ref{fig:3}b. The corresponding
spectrum (gray-shaded curve in Fig.~\ref{fig:3}a) confirms the expected large
band gap. In contrast, the $dI/dV$ spectra measured immediately at the free
zigzag edge (positions marked by red and green squares in Fig.~\ref{fig:3}b)
consistently show a filling of the entire bulk gap with spectral weight (red and
green spectra in Fig.~\ref{fig:3}a). The smooth and essentially featureless edge
spectra~\cite{footnote1} confirm a homogeneous metallic LDOS closely confined to
the zigzag edge, in full correspondence to the observations at armchair
edges~\cite{Reis2017}. Further LDOS mapping confirms that the metallic edge
channel extends along the circumference of this particular bismuthene domain,
see Fig.~S2 in the Supplementary Information. The spatial behavior of the energy-dependent LDOS along
the white-dashed path in Fig.~\ref{fig:3}b is shown in Fig.~\ref{fig:3}c.
The only spatial modulation that can be identified in this data is the
$2a_{\text{Bi}}$-periodicity due to the atomic structure of the zigzag
edge. Its purely structural origin is corroborated by its energy independence.
No additional and, in particular, energy-dependent modulations are seen that
would reflect any QPI. Consequently, we are led to conclude that no
backscattering occurs at the kinks between the piecewise straight edge segments,
consistent with the topological protection of the helical edge states at the
free zigzag edges.


Completely different behavior is observed at the DBs. Figure~\ref{fig:3}e
depicts a DB separating two bismuthene phase domains in the upper and lower half
of the image. Like the free zigzag edge, the DB is not straight
throughout, but consists of three adjacent straight segments interrupted from
each other by kinks (strictly speaking: double-kinks). The local tunneling
spectra are measured off and on the DB (positions marked by the colored
squares in Fig.~\ref{fig:3}e) and shown in Fig.~\ref{fig:3}d. While the bulk
spectrum (gray) is practically unchanged with respect to the previous case, the
DB spectra (green, blue, and red curves) differ significantly from the smooth
gap filling seen for free zigzag edges. They are instead characterized by rapid
variations with energy evolving into distinct LDOS peaks near and below zero
bias, (labeled $E_1$ -- $E_4$ in Fig.~\ref{fig:3}d). The peak intensities display a
strong spatial modulation along the DB, even when measured at topographically
equivalent positions (see, e.g., green vs. blue spectrum), pointing towards a
non-structural origin. This is even better seen in Fig.~\ref{fig:3}f showing the
full energy-dependent LDOS measured along the DB. The spatial $dI/dV(E,x)$
dependence not only reflects the DB's morphological periodicity of
$2a_{\text{Bi}}$, but, in stark contrast to the free zigzag edge, features
additional periodic modulations at the discrete energies $E_1$ to $E_4$ of the
spectral peaks in Fig.~\ref{fig:3}d. Their wavelength decreases towards higher
energy, reminiscent of standing waves in a resonator. This interpretation is
further corroborated by the two-dimensional LDOS maps of the DB 
in Fig.~\ref{fig:3}g measured at the resonance energies $E_1$ to $E_4$.
The modulations are indeed wave-like along the direction of the DB, with their
wavelength $\lambda$ directly inferred from the width-integrated $dI/dV$ line
profiles also contained in Fig.~\ref{fig:3}g. All observations are consistent with
the formation of electronic Fabry-P\'{e}rot states~\cite{Seo2010}, resulting
from QPI due to backscattering off the DB kinks. We note in passing that at 
higher energies (bias $> \, E_4$) the QPI wavelength will become comparable to 
the lattice constant, resulting in complex beating between
structural and electronic periodicities. This may account for the complicated
$dI/dV$ spectral behavior in this energy region (see Fig.~\ref{fig:3}d).

\begin{figure*}
	\centering
	\includegraphics[width = \textwidth ]{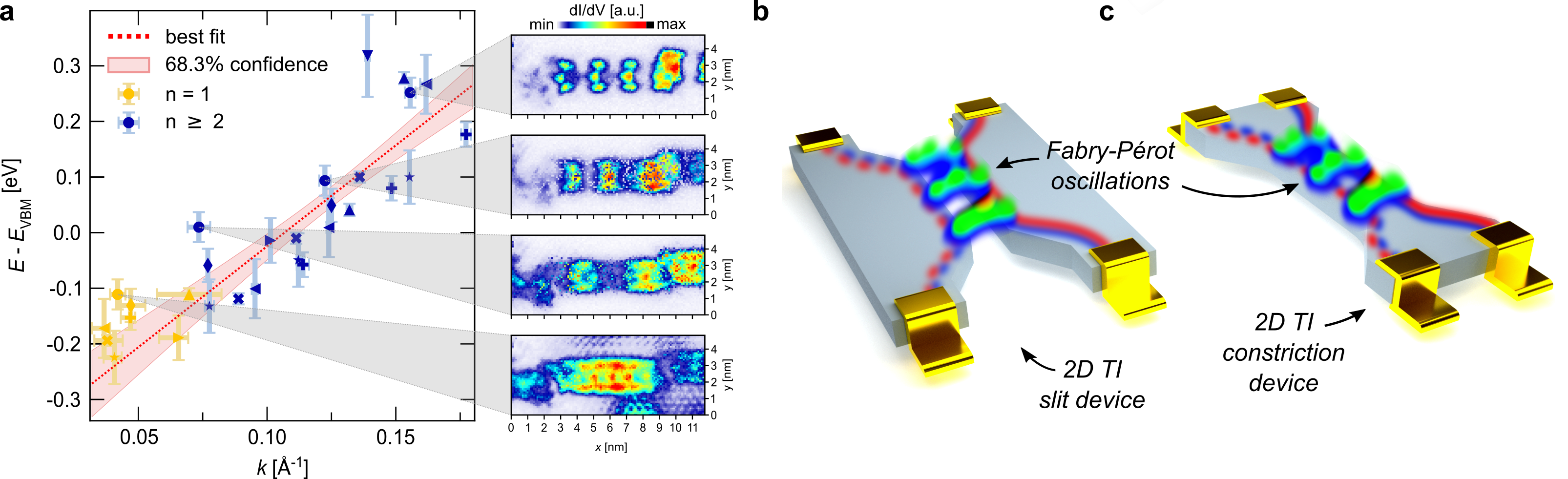}
	\caption{
	\textbf{Fabry-P\'{e}rot interferometry in a 2D TI nanoconstriction.} 
	\textbf{a}, 
		Fabry-P\'{e}rot resonator modes labeled by index $n$ as a function of $k =
		\pi/\lambda$. The modes of equal DBs are assigned equal marker symbols. Red
		dotted line: Least squares fit to the data for $n \geq 2$ (see Methods for
		details). Red shaded region: 68.3$\%$ confidence bands of the fit.
		The constant energy $dI/dV$ maps on the right-hand side (taken from Fig.~\ref{fig:3}g)
		are assigned to the respective data points. The energy error bars reflect the
		s.d. related to the valence band onset ($E_{\text{VB}}$) determination, see
		Fig.~S4 in the Supplementary Information. The $k$ error bars for $n \geq 2$ relate to the s.d. in
		determining $\lambda$. The $k$ error bars for $n = 1$ relate to the estimated
		error in determining the length of the specific DB.
	\textbf{b, c}, 
		Schematic four-terminal devices for electronic Fabry-P\'{e}rot interferometry based on a QSH insulator in slit or, alternatively,  constriction geometry.
	}
	\label{fig:4}
\end{figure*}

Our combined data thus provide clear evidence that the protection against
backscattering observed for the free zigzag (and armchair~\cite{Stuehler2020})
edge is lifted for the metallic DB states, at least partially. What remains to
be shown is that the Fabry-P\'{e}rot states in the DBs are composed of
coupled topological edge states and not just caused by QPI of trivial boundary
states. For this purpose we utilize the fact that the standing wave excitations
carry information on the $E(k)$ dispersion of the underlying electronic states.
It can be directly inferred from our STS results, with a first hint already
given by the $dI/dV(E,x)$ data in Fig.~\ref{fig:3}f. Their Fourier transform
into reciprocal space is shown in Fig.~\ref{fig:3}h where the Fabry-P\'{e}rot
modulations generate a linear $E(k)$ structure (marked by the yellow line),
as expected for a Dirac-like topological edge state. For a more systematic
determination of the dispersion we have analyzed the standing wave excitations 
in the $dI/dV$ maps of many different DB straight and defect free sections, with lengths 
ranging from \SI{3.2}{\nano\metre} to \SI{8.6}{\nano\metre}. For each DB section 
we determined the wavelength $\lambda$ of its standing waves as a function of 
energy. In order to correct for extrinsic energy shifts due to local variations of the 
chemical potential (see Supplementary Information), all energies are referred 
to the bulk valence band onset $E_\text{VBM}$, individually measured in the
local vicinity of each respective DB. The standing wave energies thus obtained are plotted in 
Fig.~\ref{fig:4}a as function of $k = \pi/\lambda$. The data points are consistently described by a linear dispersion of the form
\begin{equation}
E-E_{\text{VB}} = E_\text{D} + \hbar v_\text{F} k,
\label{eq:1} 
\end{equation} 
where $E_\text{D}$ is the Dirac point energy relative to $E_\text{VBM}$ and
$v_\text{F}$ the Fermi velocity. A possible small hybridization gap due to
inter-edge tunneling (cf.~Fig.~\ref{fig:1}b) is not resolved here. A numerical
fit of our data with the dispersion of equation~(\ref{eq:1}) yields $E_\text{D} =
\SI{-0.39 \pm 0.08}{\electronvolt}$ and $\hbar v_\text{F} = \SI{3.6 \pm
0.6}{\electronvolt \angstrom}$. For comparison, the theoretical Fermi velocity
predicted for the bismuthene zigzag edge is $\hbar v_\text{F} =
\SI{4.3}{\electronvolt \angstrom}$~\cite{footnote3}. This good correspondence
provides smoking gun evidence that the Fabry-P\'{e}rot states observed in the
bismuthene DBs are indeed derived from pairs of topological edge states, and that
their coupling opens a channel for single-particle backscattering,
i.e., that their topological protection is lifted.


It is interesting to put these results into perspective with respect to related
recent work. Pedramrazi \textit{et al.}~\cite{Pedramrazi2019} have conducted
STM/STS on similar DBs in the 2D TI $1T'$-WSe$_{2}$. While they report
qualitative differences in spectral line shape in the DBs versus interfaces
between topological and trivial phases, no backscattering-induced QPI could be
detected. Further, Howard \textit{et al.}~\cite{Howard2021} do observe QPI
at step edges of the quantum anomalous Hall material
Co$_{3}$Sn$_{2}$S$_{2}$ induced by random local impurities. 
They relate their findings to the coupling of chiral edge states in this 
material system with broken time-reversal symmetry.


In summary, the intrinsic phase-slip DBs in bismuthene realize the direct
coupling between spin-momentum locked edge states on either side of the DB. This
results in the loss of their topological protection against single-particle
back-scattering, manifesting itself in QPI, i.e., the formation of
Fabry-P\'{e}rot states. By reversing the argument, the absence of any QPI
signatures in the free zigzag edges provides strong experimental confirmation
that the metallic edge states in the bismuthene band gap are protected against
scattering and hence indeed of topological character.
These findings have bearing for quantum transport experiments on bismuthene. 
If not performed on a single phase domain, the presence of DBs may introduce unwanted dissipation and
hence impede the detection of a clean QSH effect. From a broader perspective,
our direct wave function mapping of coupled topological edges strongly
encourages novel device concepts for 2D TI-based electronic Fabry-P\'{e}rot
interferometry (FPI). Previous experimental realizations of electronic FPI have
their drawbacks, such as lack of scalability when using carbon
nanotubes~\cite{Liang2001}, or the need for high magnetic fields when based on
chiral edge states in a quantum Hall device. Instead, if narrow enough a simple slit or constriction nanopatterned into a QSH insulator (see Fig.~\ref{fig:4}b, c) that is suitable for this lithographic process
is sufficient to generate coupling between its helical edge states, similar to the situation in our bismuthene DBs. Additionally, suitable
gate electrodes could be used to tune the effective length and edge coupling
strength of the constriction and hence provide wide control of its electronic
transmission. Side gates have also been suggested to allow separate control of
charge and spin transport~\cite{Sternativo2014, Krueckl2011, Romeo2012}. We are
confident that our microscopic study of edge coupling in a QSH insulator will
fertilize the realization of such quantum transport experiments.



{\noindent
	\textbf{Methods}
}
\begin{scriptsize}
Bismuthene was grown on n-doped 4H-SiC(0001) substrates with a resistivity of
$\SI{0.01}{\ohm \cm} - \SI{0.03}{\ohm \cm}$ at room-temperature. The dopant
concentration of the substrate is $\SI{5e18}{\per \cubic\centi\meter} -
\SI{1e19}{\per \cubic\centi\meter}$. The STM/STS measurements have been
performed with a commercial low-temperature STM from Scienta Omicron GmbH under UHV
conditions ($p_\text{base} = \SI{2e-11}{\milli \bar}$). Topographic STM images
are recorded as constant current images. After stabilizing the tip at a voltage
and current set-point $V_\text{set}$ and $I_\text{set}$, respectively, the
feedback loop is opened and STS spectra are obtained making additional use of a
standard lock-in technique with a modulation
amplitude between $V_\text{mod}=\SI{5}{\milli \volt}$ and $V_\text{mod}=\SI{12}{\milli \volt}$. 
All measurements were performed at liquid He temperature, i.e., $T = \SI{4.35}{\kelvin}$. Because of
external modulation the instrumental resolution results in a Gaussian energy
broadening (FWHM: $\delta \epsilon = 2.5 e V_\text{mod}$) in addition to the
thermal broadening. Prior to every measurement on bismuthene we assured that the
tip DOS is metallic by a reference measurement on a Ag(111) surface.\\
The linear fit of the energy vs. momentum data in Fig.~\ref{fig:4}a was
performed for data points with level index $n \geq 2$ only. For such modes the
wavelength $\lambda$ of the electronic standing waves is easily determined from the
separation of two adjacent charge maxima (see Fig.~\ref{fig:3}g). In contrast,
for $n = 1$ modes $\lambda$ has to be estimated from the distance of the outer
nodal points, i.e., the longitudinal borders of the relevant straight DB section which
experimentally are less precisely identified.\\
\end{scriptsize} 


{\noindent
	\textbf{Acknowledgements}
}
\begin{scriptsize}
	The authors are indebted to Werner Hanke for enlightening discussions and to Johannes Weis for helpful technical assistance.
	This work was supported by the Deutsche Forschungsgemeinschaft
	(DFG) through the W\"urzburg-Dresden Cluster of Excellence on Complexity and Topology in Quantum Matter - \textit{ct.qmat} (EXC 2147, project-id 39085490) and the Collaborative Research Center SFB 1170 "ToCoTronics" (project-id 258499086). \\
\end{scriptsize}

\bibliographystyle{naturemag}

\end{document}